\documentclass[sigconf]{acmart}
\AtBeginDocument{%
  \providecommand\BibTeX{{%
    \normalfont B\kern-0.5em{\scshape i\kern-0.25em b}\kern-0.8em\TeX}}}

\copyrightyear{2024}
\acmYear{2024}
\setcopyright{acmlicensed}\acmConference[RecSys '24]{18th ACM Conference on Recommender Systems}{October 14--18, 2024}{Bari, Italy}
\acmBooktitle{18th ACM Conference on Recommender Systems (RecSys '24), October 14--18, 2024, Bari, Italy}
\acmDOI{10.1145/3640457.3688106}
\acmISBN{979-8-4007-0505-2/24/10}

\usepackage{booktabs}
\usepackage{multirow}
\usepackage{algorithm}
\usepackage{algorithmic}
\usepackage{enumitem}
\usepackage{xcolor}
\usepackage{subfigure}

\newcommand{\eg}{\emph{e.g.}}
\newcommand{\ie}{\emph{i.e.}}

\begin{document}


\title{FLIP: Fine-grained Alignment between ID-based Models and Pretrained Language Models for CTR Prediction}

\author{Hangyu Wang}
\authornote{Both authors contributed equally to this research.}
\email{hangyuwang@sjtu.edu.cn}
\affiliation{%
  \institution{Shanghai Jiao Tong University}
  \country{Shanghai, China}
}

\author{Jianghao Lin}
\authornotemark[1]
\email{chiangel@sjtu.edu.cn}
\affiliation{%
  \institution{Shanghai Jiao Tong University}
  \country{Shanghai, China}
}

\author{Xiangyang Li}
\email{lixiangyang34@huawei.com}
\affiliation{%
  \institution{Huawei Noah’s Ark Lab}
  \country{Shenzhen, China}
}

\author{Bo Chen}
\email{chenbo116@huawei.com}
\affiliation{%
  \institution{Huawei Noah’s Ark Lab}
  \country{Shanghai, China}
}

\author{Chenxu Zhu}
\email{zhuchenxu1@huawei.com}
\affiliation{%
  \institution{Huawei Noah’s Ark Lab}
  \country{Shanghai, China}
}

\author{Ruiming Tang}
\email{tangruiming@huawei.com}
\affiliation{%
  \institution{Huawei Noah’s Ark Lab}
  \country{Shenzhen, China}
}

\author{Weinan Zhang}
\authornote{Corresponding author.}
\email{wnzhang@sjtu.edu.cn}
\affiliation{%
  \institution{Shanghai Jiao Tong University}
  \country{Shanghai, China}
}

\author{Yong Yu}
\email{yyu@sjtu.edu.cn}
\affiliation{%
  \institution{Shanghai Jiao Tong University}
  \country{Shanghai, China}
}


\renewcommand{\shortauthors}{Hangyu Wang et al.}

\begin{abstract}

Click-through rate (CTR) prediction plays as a core function module in various personalized online services. 
The traditional ID-based models for CTR prediction take as inputs the one-hot encoded ID features of \emph{tabular modality}, which capture the collaborative signals via feature interaction modeling. 
But the one-hot encoding discards the semantic information included in the textual features. 
Recently, the emergence of Pretrained Language Models (PLMs) has given rise to another paradigm, which takes as inputs the sentences of \emph{textual modality} obtained by hard prompt templates and adopts PLMs to extract the semantic knowledge. 
However, PLMs often face challenges in capturing field-wise collaborative signals and distinguishing features with subtle textual differences.
In this paper, to leverage the benefits of both paradigms and meanwhile overcome their limitations, we propose to conduct \underline{F}ine-grained feature-level A\underline{L}ignment between \underline{I}D-based Models and \underline{P}retrained Language Models (FLIP) for CTR prediction. 
Unlike most methods that solely rely on global views through instance-level contrastive learning, we design a novel jointly masked tabular/language modeling task to learn fine-grained alignment between tabular IDs and word tokens.
Specifically, the masked data of one modality (\ie, IDs and tokens) has to be recovered with the help of the other modality, which establishes the feature-level interaction and alignment via sufficient mutual information extraction between dual modalities. 
Moreover, we propose to jointly finetune the ID-based model and PLM by adaptively combining the output of both models, thus achieving superior performance in downstream CTR prediction tasks.
Extensive experiments on three real-world datasets demonstrate that FLIP outperforms SOTA baselines, and is highly compatible with various ID-based models and PLMs. 
The code is available\footnote{PyTorch version: \url{https://github.com/justarter/FLIP}.}\footnote{MindSpore version: \url{https://github.com/mindspore-lab/models/tree/master/research/huawei- noah/FLIP}.}.

\end{abstract}

\ccsdesc[500]{Information systems~Recommender systems}
\keywords{Fine-grained Alignment, Pretrained Language Model, CTR prediction, Recommender Systems}


\maketitle

\section{Introduction}

Click-through rate (CTR) prediction is the core component of various personalized online services (\eg, web search~\cite{lin2021graph,fu2023f,dai2021adversarial}, recommender systems~\cite{xi2023bird,DeepFM}). 
It aims to estimate a user's click probability towards each target item, given a particular context~\cite{zhang2021deep,lin2023map,xi2023towards}.
Recently, the emergence of Pretrained Language Models (PLMs)~\cite{qiu2020pre} has facilitated the acquisition of extensive knowledge and enhanced reasoning abilities, hence introducing a novel paradigm for predicting CTR directly through natural language~\cite{lin2023can}.

On the one hand, the traditional \textbf{ID-based models} for CTR prediction adopt the one-hot encoding to convert the input data into ID features, which we refer to as \textbf{tabular data modality}. An example is shown as follows:
\begin{equation}
\underbrace{[0,0,1,\cdots,0]}_{\texttt{UserID=02}}\ \underbrace{[0,1]}_{\texttt{Gender=Male}}\ \underbrace{[0,1,\cdots,0]}_{\texttt{ItemID=01}}\ \underbrace{[0,1,\cdots,0]}_{\texttt{Genre=Action}}\ 
    \label{eq:onehot}
\end{equation}
Deriving from POLY2~\cite{POLY2} and FM~\cite{FM}, the key idea of these models is to capture the complex high-order feature interaction patterns across multiple fields by different operators (\eg, product~\cite{PNN,DCNv2}, convolution~\cite{cfm,FGCNN}, and attention~\cite{AutoInt,AFM}).
On the other hand, the development of PLMs has ushered in another modeling paradigm that utilizes the \textbf{PLM} as a text encoder or recommender directly.
The input data is first transformed into textual sentences via hard prompt templates, which could be referred to as \textbf{textual data modality}. 
In this way, the semantic information among original data is well preserved as natural languages (\eg, gender feature as text ``male'' instead of ID code ``[0,1]'').
Then, PLMs~\cite{devlin2018bert,liu2019roberta} encode and understand the textual sentences, and thus turn CTR prediction into either a binary text classification problem~\cite{liu2022ptab,CTRBERT} or a sequence-to-sequence task~\cite{P5,geng2023vip5}. However, both of the above models (\ie, ID-based and PLMs) possess inherent limitations.

\begin{figure}
    \centering
    \includegraphics[width=0.47\textwidth]{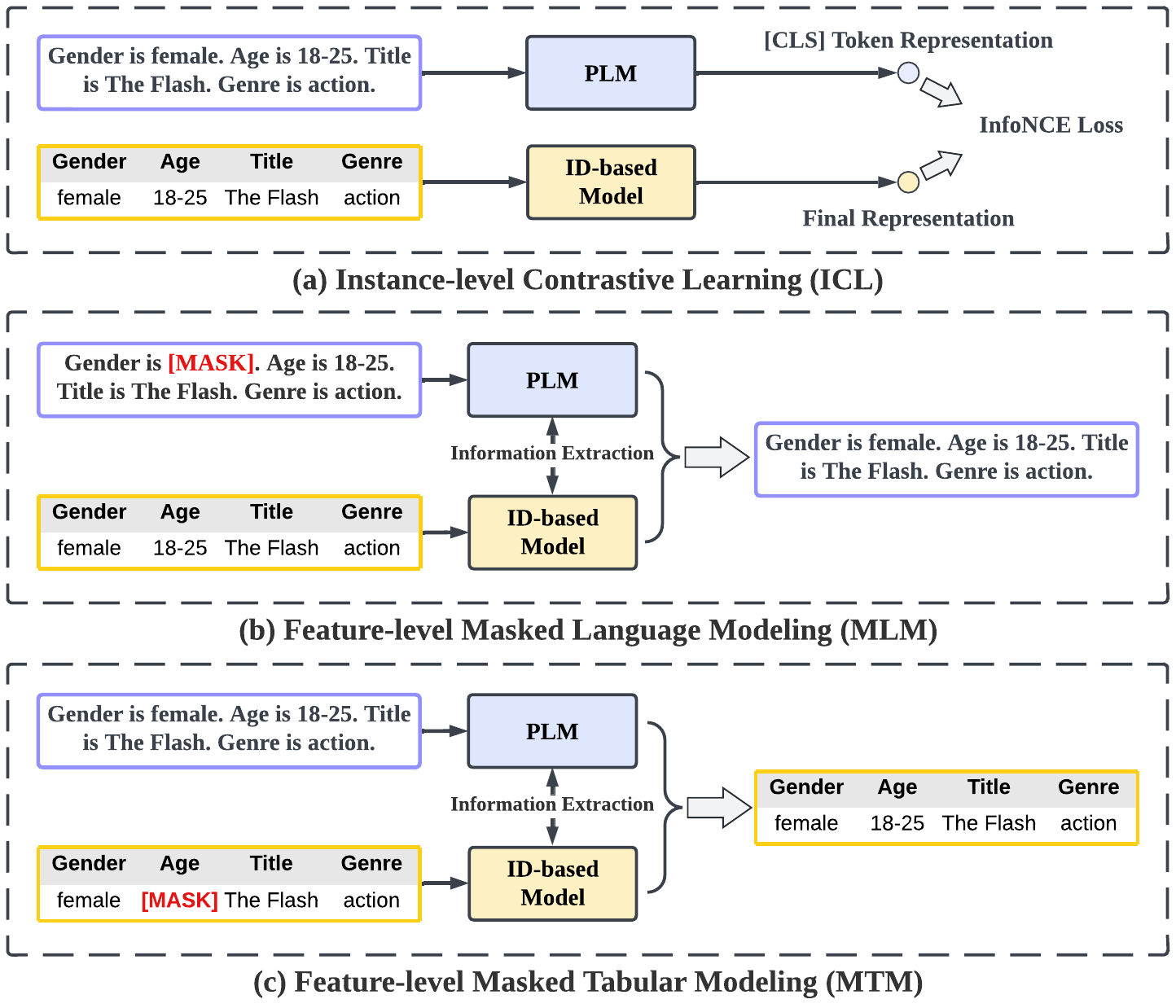}
    \vspace{-10pt}
    \caption{Three cross-modal pretraining tasks. Task (a) provides coarse-grained instance-level alignment via contrastive learning, while tasks (b) and (c) achieve fine-grained feature-level alignment through jointly masked modality modeling.}
    \vspace{-15pt}
    \label{fig:illustration}
\end{figure}

ID-based models utilize the one-hot encoding that discards the semantic information included in the textual features and fails to capture the semantic correlation between feature descriptions~\cite{li2023ctrl,lin2023clickprompt}. In addition, ID-based models rely heavily on the user interactions and may struggle in scenarios with sparse interactions~\cite{schein2002methods}. The aforementioned issues can be effectively alleviated by PLMs. PLMs excel in understanding the context and meaning behind textual features, and use their knowledge and reasoning capabilities to achieve robust performance in sparse-interaction situations~\cite{zhao2023survey, wei2023llmrec}.

Furthermore, PLMs also have certain limitations. 
PLMs struggle to understand field-wise collaborative signals because their input data is formulated as textual sentences, which are broken down into subword tokens, thus splitting the field-wise features~\cite{ren2023representation,bao2023tallrec}. 
Additionally, PLMs may fail to recognize subtle differences between different feature descriptions~\cite{clip,shi2022emscore} because it is hard to distinguish features that exhibit little textual variations in natural language~\cite{zhang2023bridging,harte2023leveraging} (\eg, in terms of the movie name, ``The Room'' and ``Room'' are two very similar movies literally).
Fortunately, the ID-based model can perceive field-wise collaborative signals with various model structures and distinguish each different feature with the distinctive ID encodings.

To this end, it is natural to bridge two paradigms to leverage the benefits of both modalities while overcoming their limitations.
Moreover, the fine-grained feature-level alignment between tabular IDs and word tokens is critical, enabling ID-based models to perceive semantic information corresponding to each feature ID and allowing PLMs to clearly distinguish features with similar text but different IDs. 
However, most existing methods~\cite{li2023ctrl,ren2023representation} align both modalities by instance-level contrastive learning, as shown in Figure~\ref{fig:illustration}(a). 
They only rely on the global view and lack supervision to encourage fine-grained alignment between IDs and tokens, potentially causing representation degeneration problems~\cite{qiu2022contrastive,gao2019representation,li2020sentence}.


In this paper, we propose to conduct \underline{F}ine-grained Feature-level A\underline{L}ignment between \underline{I}D-based Models and \underline{P}retrained Language Models (\textbf{FLIP}) for CTR prediction. 
FLIP is a model-agnostic framework that adopts the common pretrain-finetune scheme~\cite{lin2023map,devlin2018bert}.
For \emph{pretraining} objectives, as illustrated in Figure~\ref{fig:illustration}(b) and \ref{fig:illustration}(c), we propose to build jointly masked tabular and language modeling, where the masked features (\ie, IDs or tokens of specific features) of one modality are recovered with the help of another modality. 
This is motivated by the fact that both tabular and textual data convey almost the same information of the original raw data, but only in different formats.
In order to accomplish the masked feature reconstruction, each single model (ID-based model or PLM) is required to seek and exploit the valuable knowledge embedded in the other model that corresponds to the masked features, thereby achieving fine-grained feature-level cross-modal interactions.
Then, as for \emph{finetuning}, we propose a simple yet effective adaptive finetuning approach to combine the predictions from both models for downstream CTR estimation.

The main contributions of this paper are as follows:
\begin{itemize}[leftmargin=10pt]
    \item We highlight the importance of utilizing fine-grained feature-level alignment between tabular IDs and word tokens, in order to explore the potential of enhancing the performance of existing recommender systems.
    \item We propose a model-agnostic framework FLIP, where the jointly masked tabular and language modeling tasks are involved. In these tasks, the masked features of one modality have to be recovered with the help of another modality, thus learning fine-grained feature-level cross-modal interactions. FLIP also employs an adaptive finetuning approach to combine the predictions from the ID-based model and PLM for improving performance.
    \item Extensive experiments on three real-world public datasets demonstrate the superiority of FLIP, compared with existing baseline models. Moreover, we validate the model compatibility of FLIP in terms of both ID-based models and PLMs.
\end{itemize}

\section{Preliminaries}
\label{sec:preliminary}

\subsection{ID-based Models for CTR Prediction}
The traditional CTR prediction is modeled as a binary classification task~\cite{DeepFM, PNN}, whose dataset is presented as $\{(\textbf{x}_i,y_i)\}_{i=1}^N$, where $y_i \in \{1,0\}$ is the label indicating user's actual click behavior, and $\textbf{x}_i = [x_{i,1},x_{i,2},\ldots,x_{i,F}]$ is categorical tabular data with $F$ different fields.
The goal of CTR prediction is to estimate the click probability $P(y_i=1|\textbf{x}_i)$ based on input $\textbf{x}_i$. 

Generally, ID-based models adopt the ``Embedding \& Feature Interaction'' paradigm~\cite{AutoInt,xDeepFM}: (1) the input $\textbf{x}_i$ is first transformed into one-hot vectors, which are then mapped to low-dimensional embeddings via an embedding layer. (2) Next, Feature Interaction (FI) Layer is used to process the embeddings, compute complex feature interactions and generate a dense representation $\textbf{v}_i$. (3) Finally, the prediction layer (usually a MLP module) estimates the click probability $\hat{y_i} \in [0,1]$ based on the dense representation $\textbf{v}_i$. The ID-based model is trained with the binary cross-entropy (BCE) loss in an end-to-end manner:
\begin{equation}
    \mathcal{L}_{BCE}(y,\hat{y})= - \frac{1}{N}\sum\nolimits_{i=1}^N \left[y_i \log \left( \hat{y_i} \right)+(1-y_i) \log \left(1-\hat{y_i}\right) \right]
\end{equation}

\subsection{PLMs for CTR Prediction}
As PLMs have shown remarkable success in a wide range of tasks due to their extensive knowledge and reasoning capabilities~\cite{liu2019roberta,brown2020language}, researchers now tend to leverage their proficiency in natural language understanding and world knowledge modeling to solve the CTR prediction task~\cite{lin2023can,liu2023pre}. 

Different from ID-based models, (1) PLMs first transform the input data $\textbf{x}_i$ into the textual sentence $\textbf{x}^{text}_i$ via hard prompt templates. (2) Then the textual sentence $\textbf{x}^{text}_i$ is converted into meaningful lexical tokens through tokenization, which are embedded into a vector space. (3) Next, PLMs utilize their transformer-based neural networks to process the token embeddings obtained from the previous step, generating the contextually coherent representation $\textbf{w}_i$. (4) Finally, we can either add a randomly initialized classification head (usually MLP) on the representation $\textbf{w}_i$ to perform binary classification and predict the click label $y_i\in \{0,1\}$~\cite{liu2022ptab,li2023ctrl}, or add a language modeling head to do causal language modeling tasks and predict the likelihood of generating the next keyword (\eg, ``yes'' or ``no'') through a verbalizer~\cite{P5,bao2023tallrec}.
\section{Methodology}
\label{sec:method}

\subsection{Overview of FLIP}

\begin{figure*}
    \centering
    \includegraphics[width=\linewidth]{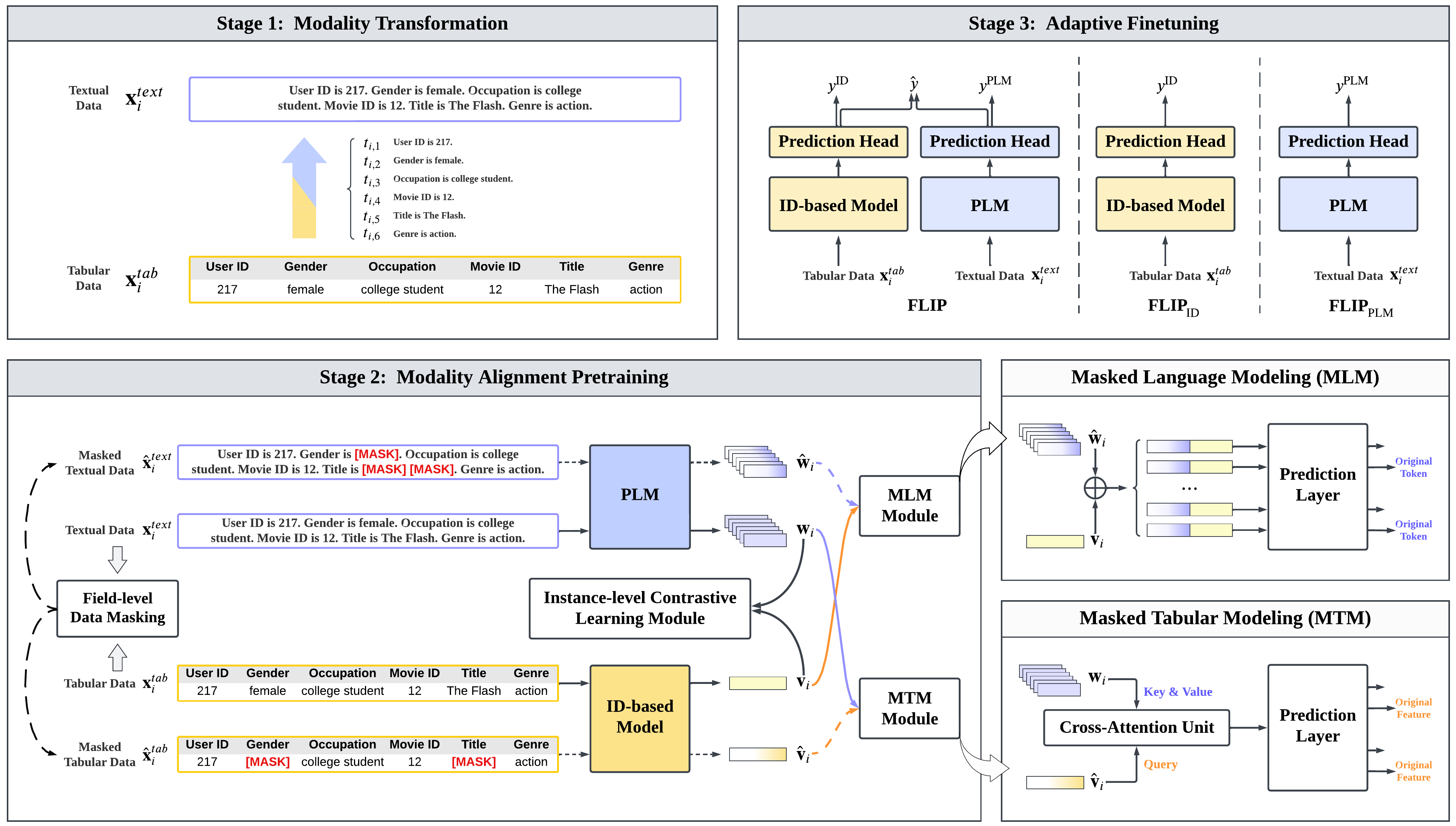}
    \vspace{-15pt}
    \caption{The overall framework of our proposed FLIP. 
    }
    \vspace{-15pt}
    \label{fig:modality alignment}
\end{figure*}

Figure~\ref{fig:modality alignment} depicts the architecture of FLIP, which consists of three stages: \textbf{modality transformation}, \textbf{modality alignment pretraining} and \textbf{adaptive finetuning}.
Firstly, FLIP transforms the raw data from tabular modality into textual modality.
Then, in the modality alignment pretraining, we employ the jointly masked language/tabular modeling task to learn fine-grained modality alignments.
Lastly, we propose a simple yet effective adaptive finetuning strategy to further enhance the performance on CTR prediction.

Hereinafter, we omit the detailed structure of PLMs and ID-based models, since FLIP serves as a model-agnostic framework and is compatible with various backbone models.

\subsection{Modality Transformation}
\label{sec:Modality Transformation}

Standard PLMs take sequences of words as inputs~\cite{devlin2018bert,radford2018improving}. 
Consequently, the modality transformation aims to convert the tabular data $\textbf{x}_i^{tab}$ into textual data $\textbf{x}_i^{text}$ via hard prompt templates.
Previous works~\cite{li2023ctrl,liu2022ptab,hegselmann2023tabllm} have suggested that sophisticated templates (\eg, with more vivid descriptions) for textual data construction might mislead the PLM and make it fail to grasp the key information in the texts.
Hence, we adopt the following simple yet effective transformation template:
\begin{equation}
\begin{aligned}
    t_{i,f}&=[m_f\oplus \text{``}is\text{''}\oplus v_{i,f}\oplus \text{``}.\text{''}], \;f\in\{1, \cdots, F\} \\
    \textbf{x}^{text}_i&=[t_{i,1}\oplus t_{i,2}\oplus \cdots \oplus t_{i,F}],
\end{aligned}
\end{equation}
where $m_f$ is the name of $f$-th field (\eg, \texttt{gender}), $v_{i,f}$ denotes the feature value of $f$-th field for input $x_i^{tab}$ (\eg, \texttt{female}), and $\oplus$ indicates the concatenation operator. 

An illustrative example is given in Figure~\ref{fig:modality alignment}~(Stage 1), where we first construct a descriptive sentence $t_{i,f}$ for each feature field, and then concatenate them to obtain the final textual data. In this way, we preserve the semantic knowledge of both field names and feature values with minimal preprocessing and no information loss. Both tabular data and textual data can be considered to contain almost the same information of raw data, albeit in different modalities.

\subsection{Modality Alignment Pretraining}
\label{sec:Modality Alignment Pretraining}

As shown in Figure~\ref{fig:modality alignment}~(Stage 2), after obtaining the paired text-tabular data $(\textbf{x}_i^{text},\textbf{x}_i^{tab})$ from the same raw input, we first perform field-level data masking to obtain the corrupted version of input pair, $(\hat{\textbf{x}}_i^{text},\hat{\textbf{x}}_i^{tab})$.
Then, the PLM $h_{\text{PLM}}$ and ID-based model $h_{\text{ID}}$ encode the input pair to obtain the dense representations $(\textbf{w}_i,\hat{\textbf{w}}_i)$ and $(\textbf{v}_i,\hat{\textbf{v}}_i)$ for textual and tabular modalities, respectively.
Next, we apply three different pretraining objectives to achieve both feature-level and instance-level alignments between PLMs and ID-based models: 
\begin{itemize}[leftmargin=10pt]
    \item \emph{Feature-level Masked Language Modeling} (MLM) requires the model to recover the original tokens from the corrupted textual context with the help of complete tabular data.
    \item \emph{Feature-level Masked Tabular Modeling} (MTM) requires the model to recover the original feature IDs from the corrupted tabular context with the help of complete textual data.
    \item \emph{Instance-level Contrastive Learning} (ICL) draws positive samples together (\ie, different modalities of the same input) and pushes apart negative sample pairs.
\end{itemize}
The jointly masked modality modeling task learns fine-grained modality interactions by the way of mask-and-predict, while ICL aligns the modalities from the perspective of the global consistency.

\subsubsection{\textbf{Field-level Data Masking}}
\label{sec:method field level data masking}

We first perform the field-level data masking strategy for textual and tabular data, respectively.

For \emph{textual data}, we propose to mask tokens at the field level.
We first randomly select a certain ratio $r_{text}$ of the feature fields to be corrupted, and then mask all the consecutive tokens that constitute the entire text of the corresponding feature value. We denote the index set of masked tokens as $\mathcal{I}^{text}$. 
Note that this is quite different from the random token-level masking strategy for common language models (\eg, BERT~\cite{devlin2018bert}).

For example, suppose the sentence tokenization from \texttt{occupation} field is [``occupation'', ``is'', ``college'', ``student'']. 
The outcome of field-level masking should be [``occupation'', ``is'', \texttt{[MASK]}, \texttt{[MASK]}]. 
But the outcome of token-level masking might be [\texttt{[MASK]}, ``is'', ``college'', ``student''] or [``occupation'', ``is'', ``college'', \texttt{[MASK]}]. Obviously, token-level masked tokens can be easily inferred solely based on the textual context, thus limiting the cross-modal interactions and losing the generalization ability~\cite{liu2021self}. 

For \emph{tabular data}, following previous works~\cite{lin2023map,wang2020masked}, we uniformly sample a certain ratio $r_{tab}$ of the fields, and then replace the corresponding feature with an additional \texttt{<MASK>} feature, which is set as a special feature in the embedding table of ID-based model. 
The \texttt{<MASK>} feature is not field-specific, but shared by all feature fields to prevent introducing prior knowledge about the masked field~\cite{lin2023map}.
Similarly, we denote the index set of masked fields as $\mathcal{I}^{tab}$. 

After the field-level data masking, we obtain the corrupted version for both modalities, \ie, $(\hat{\textbf{x}}_i^{text},\hat{\textbf{x}}_i^{tab})$. It is worth noting that the selected fields to be masked  do not necessarily have to be the same for textual and tabular modalities, since they serve as two independent parallel workflows.

\subsubsection{\textbf{Data Encoding}}

We employ the PLM $h_{\text{PLM}}$ and ID-based model $h_{\text{ID}}$ to encode the input pairs from textual and tabular modalities, respectively:
\begin{equation}
\begin{aligned}
    \textbf{w}_i&=h_{\text{PLM}}(\textbf{x}_i^{text}),\;\;\;\hat{\textbf{w}}_i=h_{\text{PLM}}(\hat{\textbf{x}}_i^{text}), \\
    \textbf{v}_i&=h_{\text{ID}}(\textbf{x}_i^{tab}),\;\;\;\;\;\;\;\hat{\textbf{v}}_i=h_{\text{ID}}(\hat{\textbf{x}}_i^{tab}). \\
\end{aligned}
\end{equation}
Here, $\textbf{w}_i=[w_{i,l}]_{l=1}^L\in\mathbb{R}^{L\times D_{text}}$ is the set of hidden states from the last layer of the PLM, where $L$ is the number of tokens for $\textbf{x}_i^{text}$ and $D_{text}$ is the hidden size of the PLM.
Note that $w_{i,1}$ is the \texttt{[CLS]} token vector that represents the overall textual input.
$\hat{\textbf{w}}_i=[\hat{w}_{i,l}]_{l=1}^L\in\mathbb{R}^{L\times D_{text}}$ also satisfies the notations above.
$\textbf{v}_i,\hat{\textbf{v}}_i\in\mathbb{R}^{D_{tab}}$ are the representations produced by ID-based model.

\subsubsection{\textbf{Masked Language Modeling}}

As shown in Figure~\ref{fig:modality alignment}~(Stage 2), the MLM module takes the text-tabular pair $(\hat{\textbf{w}}_i,\textbf{v}_i)$ as input, and attempts to recover the masked tokens.
We denote the index set of masked tokens as $\mathcal{I}^{text}$.
For each masked token with index $l\in\mathcal{I}^{text}$, we concatenate the corresponding token vector $\hat{w}_{i,l}$ with the ``reference answer'' $\textbf{v}_i$, and then feed them through the prediction layer to obtain the estimated distribution:
\begin{equation}
    q_{i,l}=g_{\text{PLM}}([\hat{w}_{i,l}\oplus \textbf{v}_i])\in\mathbb{R}^{V},
\end{equation}
where $\oplus$ is the concatenation operation, $V$ is the vocabulary size, and $g_{\text{PLM}}$ is a two-layer MLP. Finally, similar to common masked language modeling~\cite{devlin2018bert}, we leverage the cross-entropy loss for pretraining optimization:
\begin{equation}
    \mathcal{L}^{MLM}_i = \frac{1}{|\mathcal{I}^{text}|} \sum\nolimits_{l\in\mathcal{I}^{text}} \operatorname{CrossEntropy}\left(q_{i,l},\ x^{text}_{i,l}\right),
\end{equation}
where $x^{text}_{i,l}$ is the original $l$-th token.

\subsubsection{\textbf{Masked Tabular Modeling}}

Likewise, the MTM module takes the text-tabular pair $(\textbf{w}_i,\hat{\textbf{v}}_i)$ as input, and aims to recover the masked features.
To dynamically capture the essential knowledge from the corresponding tokens of the masked features, we design the cross-attention unit to aggregate the tabular representation $\hat{\textbf{v}}_i$ with the ``reference answer'' $\textbf{w}_i$:
\begin{equation}
    \textbf{u}_i = \text{Softmax}\left(\frac{\widehat{\textbf{v}}_i \textbf{Q} \textbf{w}_i^T}{\sqrt{D_{text}}}\right)\textbf{w}_i,\;\; \textbf{u}_i \in \mathbb{R}^{L}
\end{equation}
where $\textbf{Q} \in \mathbb{R}^{D_{tab}\times D_{text}}$ is the trainable cross-modal attention matrix, and $\sqrt{D_{text}}$ is the scaling factor~\cite{vaswani2017attention}.

For each masked feature with index $f\in\mathcal{I}^{tab}$, we maintain an independent MLP network $g_{\text{ID}}^{(f)}$ followed by a softmax function to compute the distribution $p_{i,f}\in \mathbb{R}^M$ over the candidate features:
\begin{equation}
\begin{aligned}
    \textbf{c}_{i,f} &= g_{\text{ID}}^{(f)}(\textbf{u}_i), \quad \textbf{c}_{i,f}\in \mathbb{R}^M, \\
    p_{i,f,j} &= \frac{\text{exp}(c_{i,f,j})}{\sum_{k=1}^M \text{exp}(c_{i,f,k})}, \quad j=1,\cdots,M,
    \label{eq:softmax}
\end{aligned}
\end{equation}
where $M$ is the size of the entire feature space. Finally, we adopt the cross-entropy loss on all the masked features:
\begin{equation}
    \mathcal{L}_i^{MTM} = \frac{1}{|\mathcal{I}^{tab}|} \sum\nolimits_{f\in\mathcal{I}^{tab}} \operatorname{CrossEntropy}(p_{i,f},\ x^{tab}_{i,f}),
\end{equation}
where $x^{tab}_{i,f}$ is the original feature of $f$-th field.

However, the loss above is actually impractical and inefficient since it has to calculate the softmax function over the entire feature space in Eq.~\ref{eq:softmax}, where $M$ is usually at million level for real-world applications. 
To this end, we adopt noise contrastive estimation (NCE)~\cite{gutmann2010noise, mikolov2013efficient, lin2023map}. 
NCE transforms a multi-class classification task into a binary classification task, where the model is required to distinguish positive features (\ie, masked features) from noise features. 
Specifically, for each masked feature $x_{i,f}^{tab}$ of $f$-th field, we sample $K$ noise features from the entire feature space according to their frequency distribution in the training set. 
Then, we utilize the binary cross-entropy (BCE) loss for MTM optimization:
\begin{equation}
    \mathcal{L}_i^{MTM} =    -\frac{1}{|\mathcal{I}^{tab}|} \sum_{f\in\mathcal{I}^{tab}} ( \log\sigma(c_{i,f,t}) + \sum_{k=1}^K \log(1-\sigma(c_{i,f,k})) ) )
\label{eq:nce}
\end{equation}
where $\sigma$ is the sigmoid function, and $t$, $k$ are the indices of the positive and noise features respectively.

\subsubsection{\textbf{Instance-level Contrastive Learning (ICL)}}

In addition to the feature-level alignment through the masked modality modeling, we also introduce contrastive learning to explicitly learn instance-level consistency between two modalities. 
The contrastive objective draws the representations of matched text-tabular pairs together and pushes apart those non-matched pairs~\cite{clip,li2021align}. 

Here, we utilize the \texttt{[CLS]} token vector $w_{i,1}$ to represent the textual input $\textbf{x}_i^{text}$.
For dimensional consistency, we employ two separate linear layers to
project the \texttt{[CLS]} token vector $w_{i,1}$ and tabular representation $\textbf{v}_i$ into $d$-dimensional vectors, $z^{text}_i$ and $z^{tab}_i$, respectively. 
Next, we adopt InfoNCE~\cite{oord2018representation} to compute the ICL loss:
\begin{equation}
\begin{split}
    \mathcal{L}^{ICL} = -\frac{1}{2B}\sum_{i=1}^B \Bigg[ &\log\left(\frac{\text{exp}( \text{sim}(z^{text}_i, z^{tab}_i) / \tau )}{\sum_{j} \text{exp}( \text{sim}(z^{text}_i, z^{tab}_j) / \tau) }  \right)  \\ + &\log\left(\frac{\text{exp}( \text{sim}(z^{tab}_i, z^{text}_i) / \tau )}{\sum_{j} \text{exp}(\text{sim}(z^{tab}_i, z^{text}_j) / \tau) }  \right) \Bigg]
\label{eq:icl}
\end{split}
\end{equation}
where $B$ is the batch size, $\tau$ is the temperature hyperparameter, and the similarity function $\text{sim}(\cdot)$ is measured by dot product.

Finally, by putting the three objectives together, the overall loss for the modality alignment pretraining stage is:
\begin{equation}
\mathcal{L}^{pretrain}=\frac{1}{B}\sum\nolimits_{i=1}^B\left(\mathcal{L}^{MLM}_i+\mathcal{L}^{MTM}_i\right)+\mathcal{L}^{ICL}.
\end{equation}


\subsection{Adaptive Finetuning}
\label{sec:Adaptive Finetuning}

After the pretraining stage, the PLM and ID-based model have learned fine-grained multimodal representations.
As depicted in Figure~\ref{fig:modality alignment}~(Stage 3), in this stage, we adaptively finetune the two models jointly on the downstream CTR prediction task with supervised click signal to achieve superior performance.

\textbf{FLIP} places a randomly initialized linear layer on the ID-based model, and another layer upon the PLM, so that these two models can output the estimated probability $\hat{y}_i^{\text{ID}}$ and $\hat{y}_i^{\text{PLM}}$ respectively. 
\begin{equation}
    \begin{aligned}
\hat{y}^{\text{ID}}_i&=\sigma(\operatorname{Linear}_{\text{ID}}(\textbf{v}_i)),\\
\hat{y}^{\text{PLM}}_i&=\sigma(\operatorname{Linear}_{\text{PLM}}(\textbf{w}_i))
    \end{aligned}
\end{equation}
And the final click probability is estimated by a weighted sum of outputs from both models:
\begin{equation}
    \hat{y}_i=\sigma\left( \alpha\times \operatorname{Linear}_{\text{ID}}(\textbf{v}_i) + (1-\alpha)\times \operatorname{Linear}_{\text{PLM}}(\textbf{w}_i) \right),
\end{equation}
where $\alpha\in$ is a learnable parameter to adaptively balance the outcomes from two models. 
To avoid performance collapse mentioned in previous works~\cite{buhlmann2012bagging,boosting}, we apply BCE objectives over the jointly estimated click probability $\hat{y}$, as well as the solely estimated click probability $\hat{y}_i^{\text{ID}}$ and $\hat{y}_i^{\text{PLM}}$ for model optimization:
\begin{equation}
    \mathcal{L}^{finetune}= \mathcal{L}_{BCE}(y,\hat y)+ \mathcal{L}_{BCE}(y,\hat{y}^{\text{ID}}) + \mathcal{L}_{BCE}(y,\hat{y}^{\text{PLM}})
\end{equation}


To delve deeper into the influence of fine-grained alignment on the single model, we define two variants: \textbf{FLIP}$_{\text{ID}}$ and \textbf{FLIP}$_{\text{PLM}}$. The former solely finetunes the ID-based model with loss $\mathcal{L}_{BCE}(y,\hat{y}^{\text{ID}})$, while the latter solely finetunes the PLM with loss $\mathcal{L}_{BCE}(y,\hat{y}^{\text{PLM}})$.

It is worth noting that FLIP is expected to achieve the superior performance since it explicitly combines the predictions from two models, while FLIP$_{\text{ID}}$ and FLIP$_{\text{PLM}}$ could more clearly reveal the effect of fine-grained alignment on a single model. 

\section{Experiment}


\subsection{Experiment Setup}

\subsubsection{Datasets}
We conduct experiments on three real-world public datasets: \href{https://grouplens.org/datasets/MovieLens/1m/}{MovieLens-1M}, \href{http://www2.informatik.uni-freiburg.de/~cziegler/BX/}{BookCrossing}, \href{https://sites.google.com/eng.ucsd.edu/ucsdbookgraph/home}{GoodReads}. 
All of the selected datasets contain user and item information. The information is unencrypted original text, thus preserving the real semantic information.
\begin{itemize}[leftmargin=8pt]
    \item \href{https://grouplens.org/datasets/MovieLens/1m/}{\textbf{MovieLens-1M}}~\cite{harper2015movielens} is a movie recommendation dataset with user-movie ratings ranging from 1 to 5. 
    Following the previous work~\cite{AutoInt,li2023ctrl}, We consider samples with ratings greater than 3 as positive, samples with ratings less than 3 as negative, and remove samples with ratings equal to 3 (\ie, neutral).
    \item \href{http://www2.informatik.uni-freiburg.de/~cziegler/BX/}{\textbf{BookCrossing}}~\cite{bookcrossing} is a book recommendation dataset and possesses user-book ratings ranging from 0 to 10. We consider samples with scores greater than 5 as positive, and the rest as negative.
    \item \href{https://sites.google.com/eng.ucsd.edu/ucsdbookgraph/home}{\textbf{GoodReads}}~\cite{goodreads1,goodreads2} is a book recommendation dataset which contains user-book ratings ranging from 1 to 5. 
    We take samples with ratings greater than 3 as positive, and the rest as negative.
\end{itemize}

Following previous works~\cite{li2023ctrl,li2022inttower}, we sort all samples in chronological order and take the first 90\% samples as the training set and the remaining as the testing set. The training set for the pretraining and finetuning stage are the same~\cite{lin2023map}. 
All our experimental results are obtained on the testing set. 
The statistics of the processed datasets are shown in Table~\ref{tab:datasets}.

\subsubsection{Evaluation Metrics}
We use commonly adopted metrics, AUC (Area Under the ROC Curve) and Logloss (binary cross-entropy loss) as the evaluation metrics. 
Notably, a slightly higher AUC or a lower Logloss (\eg, \textbf{0.001}) can be considered as a significant improvement in CTR prediction~\cite{xDeepFM,DeepFM,inttower}.

\subsubsection{Baselines}
The baseline methods can be mainly classified into three categories: (1) \textbf{ID-based} models: AFM~\cite{AFM}, PNN~\cite{PNN}, Wide\&Deep~\cite{WDL}, DCN~\cite{DCNv1}, DeepFM~\cite{DeepFM}, xDeepFM~\cite{xDeepFM}, AFN~\cite{AFN}, AutoInt~\cite{AutoInt} and DCNv2~\cite{DCNv2}, (2) \textbf{PLM-based} models: CTR-BERT~\cite{CTRBERT}, P5~\cite{P5} and PTab~\cite{liu2022ptab}, (3) \textbf{ID+PLM} models that combine ID-based model and PLM: CTRL~\cite{li2023ctrl}, MoRec~\cite{yuan2023go}. 

\begin{table}
    \vspace{-5pt}
    \caption{Statistics of processed datasets.}
    \vspace{-10pt}
    \centering
    \resizebox{.7\linewidth}{!}{
    \renewcommand\arraystretch{1.1}
    \begin{tabular}{c|cccc}
    \toprule
     Dataset   &\#Samples & \#Fields & \#Features \\ 
     \midrule
     MovieLens-1M  & 739,012 & 8 & 16,849  \\
     BookCrossing & 1,031,171 & 8 & 722,235 \\
     GoodReads & 20,122,040 & 15 & 4,565,429  \\ \bottomrule
    \end{tabular}
    }
    \vspace{-10pt}
    \label{tab:datasets}
\end{table}

\begin{table*}
\vspace{-5pt}
\caption{The overall performance of different models from three categories (\ie, ID-based, PLM-based, and ID+PLM). 
For each type of models, the best result is given in bold, and the second-best value is underlined. \emph{Rel.Impr} denotes the relative AUC improvement rate of our method against each baseline within each category. The symbol ``$\ast$'' indicates statistically significant improvement of FLIP over the best baseline with $p$-value $<$ 0.001.}
\vspace{-10pt}
\label{tab:performance}
\resizebox{0.87\textwidth}{!}{
\renewcommand\arraystretch{1.08}
\begin{tabular}{cc|ccc|ccc|ccc}
\toprule
\hline
\multicolumn{2}{c|}{\multirow{2}{*}{Model}} & \multicolumn{3}{c|}{MovieLens-1M} & \multicolumn{3}{c|}{BookCrossing} & \multicolumn{3}{c}{GoodReads} \\ 
 & & AUC  & Logloss &  Rel.Impr & AUC & Logloss & Rel.Impr & AUC & Logloss & Rel.Impr  \\ \hline
  
 \multicolumn{1}{c}{\multirow{10}{*}{ID-based}} & AFM  & 0.8449 & 0.3950  & 1.79\%  & 0.7946 & 0.5116  & 1.06\% & 0.7630 & 0.5160 & 1.96\% \\
 & PNN & 0.8546 & 0.3946   & 0.63\% & 0.7956 & 0.5131 & 0.93\% &  \underline{0.7725} & \underline{0.5055} & 0.70\%  \\  
&Wide\&Deep & 0.8509 & 0.3957 & 1.07\%  & 0.7951 & 0.5116 & 1.00\% &  0.7684 & 0.5090 & 1.24\%     \\ 
 &DCN   & 0.8509 & 0.4056 & 1.07\%  & \underline{0.7957} & 0.5108  & 0.92\%  &   0.7693 & 0.5086 & 1.12\%  \\ 
& DeepFM & 0.8539 & 0.3905  &  0.71\%  &  0.7947 & 0.5122  & 1.04\% & 0.7671 & 0.5138 & 1.41\% \\ 
  &xDeepFM  &   0.8454 & 0.3934  &  1.72\%   &  0.7953 & 0.5108  & 0.97\% & 0.7720 & 0.5079 & 0.77\% \\ 
  &AFN    & 0.8525	& \underline{0.3868}  &  0.88\%  & 0.7932 & 0.5139 & 1.24\% &  0.7654 & 0.5118 & 1.64\%  \\  
 & AutoInt &  0.8509 &	0.4013 &  1.07\%  & 0.7953 & 0.5118 & 0.97\% & 0.7716 & 0.5071 & 0.82\% \\    
 & DCNv2  &   \underline{0.8548} & 0.3893 &  0.61\%  & 0.7956 & \underline{0.5103}  & 0.93\% & 0.7724 & 0.5057 & 0.72\%  \\ 
 & \textbf{FLIP}$_{\text{ID}}$\, (Ours)  & \textbf{0.8600}* & \textbf{0.3802}* &  -  &  \textbf{0.8030}* & \textbf{0.5043}*   & - & \textbf{0.7779}* & \textbf{0.5014}*   & -  \\
  \hline \hline

 \multicolumn{1}{c}{\multirow{4}{*}{PLM-based}} & CTR-BERT   & 0.8304 & \underline{0.4131}   & 1.88\% & 0.7795 & 0.5300  & 1.65\% & 0.7385 & 0.5316  & 1.23\%   \\  
&P5  & 0.8304 & 0.4173  & 1.88\% & 0.7801 & \underline{0.5261} & 1.58\%  & 0.7365 & 0.5336  & 1.51\% \\
&PTab   & \underline{0.8426} & 0.4195 & 0.41\%  & \underline{0.7880} & 0.5384  & 0.56\% & \underline{0.7456} & \underline{0.5268} & 0.27\% \\ 
& \textbf{FLIP}$_{\text{PLM}}$ (Ours)  & \textbf{0.8460*} & \textbf{0.4127*} & - &  \textbf{0.7924*} & \textbf{0.5304*}   & - & \textbf{0.7476*} & \textbf{0.5255*}  & -  \\
\hline  \hline

 \multicolumn{1}{c}{\multirow{3}{*}{ID+PLM}}& CTRL  &  \underline{0.8572} &  \underline{0.3838}  & 0.57\%  & 0.7985 & 0.5101 & 0.95\% & \underline{0.7741} & \underline{0.5045} & 0.59\%  \\ 
& MoRec & 0.8561 & 0.3896 & 0.70\%  & \underline{0.7990} & \underline{0.5087} & 0.89\%  & 0.7731  & 0.5085  & 0.72\%  \\

& \textbf{FLIP} (Ours)&  \textbf{0.8621*}  & \textbf{0.3788*}  & -  & \textbf{0.8061*} & \textbf{0.5004*}  & - & \textbf{0.7787*} & \textbf{0.5001*}   & -    \\ \hline  \bottomrule          
\end{tabular}
}
\vspace{-5pt}
\end{table*}

\subsubsection{Implementation Details}

All of our experiments are performed on 8 NVIDIA Tesla V100 GPUs with PyTorch~\cite{paszke2019pytorch}. In the modality alignment pretraining stage, we set the text and tabular mask ratio $r_{text}$ and $r_{tab}$ both to 15\%. Unless specified otherwise, we adopt TinyBERT~\cite{jiao2020tinybert} as the PLM, and DCNv2~\cite{DCNv2} as the ID-based model. During pretraining, the model is trained for 30 epochs with the AdamW~\cite{adamw} optimizer and a batch size of 1024. The learning rate is initialized as 5e-5 followed by a cosine decay strategy. The number of noise features $K$ is 25 in Eq.~\ref{eq:nce}. The temperature $\tau$ is 0.7 in Eq.~\ref{eq:icl}. In the adaptive finetuning stage, we adopt the Adam~\cite{kingma2014adam} optimizer with learning rate selected from \{1e-5,5e-5,1e-4,5e-4,1e-3,5e-3\}. The finetuning batch size is 256 for MovieLens-1M and BookCrossing, and is 2048 for GoodReads. 

For all ID-based models in the baselines or FLIP framework, the embedding size is fixed to 32, and the size of DNN layers is [300,300,128]. 
The structure parameters and the size of PLMs in baselines are set according to their original papers. We also apply grid search to baseline methods for optimal performance.



\subsection{Performance Comparison}

We compare the recommendation performance of FLIP with three categories of baselines. We also include variants FLIP$_{\text{ID}}$ and FLIP$_{\text{PLM}}$ to reveal the effect of fine-grained alignment on a single model.
The results are shown in Table~\ref{tab:performance}, from which we can observe that:
\begin{itemize}[leftmargin=10pt]
    \item FLIP outperforms all baselines from all three categories (\ie, ID-based, PLM-based, and ID+PLM) significantly, which confirms the excellence of our proposed fine-grained feature-level alignment and adaptive finetuning approach for CTR prediction. 
    \item FLIP$_{\text{ID}}$ surpasses all ID-based baselines, while FLIP$_{\text{PLM}}$ outperforms all PLM-based baselines. These phenomena prove that fine-grained modality alignment can leverage the benefits of both models and boost their own performance.
    \item CTRL and MoRec generally outperform other baselines due to their integration of ID-based models and PLMs. However, they either overlook the fine-grained alignment between IDs and tokens or ignore cross-modal interactions, thereby degrading the performance.
    FLIP addresses these shortcomings by bridging ID-based models and PLMs through jointly masked tabular/language modeling, which enables fine-grained feature-level interactions between dual modalities, thus resulting in superior performance.
\end{itemize}

\subsection{Compatibility Analysis}

FLIP serves as a model-agnostic framework that is compatible with various backbone models.
In this section, we investigate the model compatibility in terms of different ID-based models and PLMs.

\subsubsection{Compatibility with ID-based models}

We apply FLIP to three different ID-based models, including DeepFM, AutoInt and DCNv2, while keeping TinyBERT as the PLM. The results are listed in Table~\ref{tab:differentCTR}.
We can obtain the following observations:
\begin{itemize}[leftmargin=10pt]
    \item First and foremost, FLIP$_{\text{ID}}$ consistently surpasses the corresponding vanilla ID-based model by a large margin without altering the model structure or increasing the inference cost. 
    This indicates that ID-based models of various structures can all acquire useful semantic information from PLMs through the fine-grained alignment pretraining, thereby improving performance.
    \item By jointly tuning the ID-based model and PLM, FLIP achieves the best performance across various ID-based backbone models significantly, demonstrating the superior compatibility of FLIP in terms of ID-based models. 
\end{itemize}

\begin{table}[t]
\caption{The compatibility w.r.t. different ID-based models. The PLM is fixed as TinyBERT. \emph{N/A} means to train the vanilla ID-based model from scratch. For each ID-based model, the best result is in bold, and the second-best is underlined. }
\vspace{-10pt}
\label{tab:differentCTR}
\resizebox{\linewidth}{!}{
\renewcommand\arraystretch{1.1}
\begin{tabular}{c|c|cc|cc}
\toprule
\hline
 \multicolumn{1}{c|}{\multirow{2}{*}{ID-based Model}} & \multicolumn{1}{c|}{\multirow{2}{*}{Finetuning Strategy}}  &  \multicolumn{2}{c|}{MovieLens-1M} &  \multicolumn{2}{c}{BookCrossing} \\ 
 \multicolumn{1}{c|}{} & \multicolumn{1}{c|}{} &  AUC & Logloss & AUC & Logloss  \\ \hline
 
\multicolumn{1}{c|}{\multirow{4}{*}{DeepFM}} & N/A & 0.8539 & 0.3905 & 0.7947 & 0.5122        \\
& FLIP$_{\text{ID}}$ & \underline{0.8600} & \textbf{0.3752} & \underline{0.8021} & \underline{0.5083}  \\
& FLIP$_{\text{PLM}}$ & 0.8445 & 0.4132 & 0.7892 & 0.5324  \\
& FLIP  & \textbf{0.8615} & \underline{0.3758} & \textbf{0.8033} & \textbf{0.5031}  \\ \hline
\multicolumn{1}{c|}{\multirow{4}{*}{AutoInt}} & N/A   & 0.8509 & 0.4013 & 0.7943 & 0.5118   \\
& FLIP$_{\text{ID}}$ & \underline{0.8583} & \underline{0.3827} & \underline{0.7992} & \underline{0.5092}  \\
& FLIP$_{\text{PLM}}$ & 0.8453 & 0.4126 & 0.7909 & 0.5338  \\
& FLIP  &  \textbf{0.8600} & \textbf{0.3807} & \textbf{0.8011} & \textbf{0.5050}  \\ \hline
\multicolumn{1}{c|}{\multirow{4}{*}{DCNv2}} & N/A  & 0.8548 & 0.3893 & 0.7956 & 0.5103           \\
& FLIP$_{\text{ID}}$ & \underline{0.8600} & \underline{0.3802} & \underline{0.8030} & \underline{0.5043}   \\
& FLIP$_{\text{PLM}}$ & 0.8460 & 0.4127 & 0.7924 & 0.5304   \\
& FLIP  & \textbf{0.8621} & \textbf{0.3788} & \textbf{0.8061} & \textbf{0.5004}   \\ \hline \bottomrule 
\end{tabular}
}
\vspace{-10pt}
\end{table}

\subsubsection{Compatibility with PLMs}
\label{sec:Compatibility with PLMs}

Similarly, we keep DCNv2 as the ID-based model, and select PLMs of different sizes, including TinyBERT (14.5M), RoBERTa-Base (125M)~\cite{liu2019roberta} , and RoBERTa-Large (355M)~\cite{liu2019roberta}. 
The results are in Table~\ref{tab:differentsemantic}, from which we find that:
\begin{itemize}[leftmargin=10pt]
    \item As the size of PLM grows, the performance of FLIP$_{\text{PLM}}$ continuously increases and even achieves a better AUC 0.7972 on BookCrossing with RoBERTa-Large compared with the vanilla DCNv2. 
    A larger model size would lead to larger model capacity and better language understanding ability, thus benefiting the final predictive performance.
    \item While increasing the PLM's size is expected to yield more notable performance improvements, the advantages of scaling up gradually taper off. For instance, the improvement from RoBERTa-Base to RoBERTa-large is significantly smaller than the improvement from TinyBERT to RoBERTa-Base.
    \item FLIP and FLIP$_{\text{ID}}$ outperform the DCNv2 model consistently and significantly, highlighting FLIP's ability to adapt seamlessly to different PLM sizes and architectures.
\end{itemize}

\begin{table}[t]
\caption{The compatibility w.r.t. different PLMs. The ID-based model is fixed as DCNv2. For each type of PLM, the best result is in bold, and the second-best is underlined.}
\vspace{-10pt}
\label{tab:differentsemantic}
\resizebox{\linewidth}{!}{
\renewcommand\arraystretch{1.1}
\begin{tabular}{c|c|cc|cc}
\toprule
\hline
 \multicolumn{1}{c|}{\multirow{2}{*}{PLM}} & \multicolumn{1}{c|}{\multirow{2}{*}{Finetuning Strategy}}  &  \multicolumn{2}{c|}{MovieLens-1M} &  \multicolumn{2}{c}{BookCrossing} \\ 
 \multicolumn{1}{c|}{} & \multicolumn{1}{c|}{} &  AUC & Logloss & AUC & Logloss  \\ \hline
\multicolumn{2}{c|}{DCNv2}  & 0.8548 & 0.3893 & 0.7956 & 0.5103           \\ \hline
\multicolumn{1}{c|}{\multirow{3}{*}{TinyBERT}} & FLIP$_{\text{ID}}$ & \underline{0.8600} & \underline{0.3802} & \underline{0.8030} & \underline{0.5043}   \\
& FLIP$_{\text{PLM}}$ & 0.8460 & 0.4127 & 0.7924 & 0.5304  \\
& FLIP  &  \textbf{0.8621} & \textbf{0.3788} & \textbf{0.8061} & \textbf{0.5004}    \\ \hline
\multicolumn{1}{c|}{\multirow{3}{*}{RoBERTa-Base}} & FLIP$_{\text{ID}}$ &  \underline{0.8601} & \underline{0.3784} & \underline{0.8038} & \underline{0.5030}       \\
& FLIP$_{\text{PLM}}$ & 0.8499 & 0.4053 & 0.7961 & 0.5292     \\
& FLIP  & \textbf{0.8634} & \textbf{0.3770} & \textbf{0.8083} & \textbf{0.4995}    \\ \hline
\multicolumn{1}{c|}{\multirow{3}{*}{RoBERTa-Large}} & FLIP$_{\text{ID}}$ &  \underline{0.8603} & \underline{0.3774} & \underline{0.8036} & \underline{0.5035}    \\
& FLIP$_{\text{PLM}}$ & 0.8506 & 0.4045 & 0.7972 & 0.5280       \\
& FLIP  &  \textbf{0.8650} & \textbf{0.3757} & \textbf{0.8092} & \textbf{0.4986}      \\ \hline \bottomrule       
\end{tabular}
}
\vspace{-10pt}
\end{table}

\subsection{Ablation Study}

We conduct ablation experiments for better understanding the contributions of different components in our proposed FLIP. 

\begin{table}
\caption{The results of ablation study.}
\vspace{-10pt}
\label{tab:ablation}
\resizebox{0.44\textwidth}{!}{
\renewcommand\arraystretch{1.1}
\begin{tabular}{lcccc}
\toprule
\hline
\multicolumn{1}{c}{\multirow{2}{*}{Model Variant}} & \multicolumn{2}{c}{MovieLens-1M} & \multicolumn{2}{c}{BookCrossing} \\ 
  & AUC & Logloss  & AUC & Logloss  \\ \hline
  FLIP & \textbf{0.8621} & \textbf{0.3788} & \textbf{0.8061} & \textbf{0.5004}\\
  \hline
  w/o \emph{MLM} & 0.8610 & 0.3791 & 0.8042 & 0.5035  \\
  w/o \emph{MTM} &  0.8615 & 0.3806 & 0.8053 & 0.5032 \\
  w/o \emph{ICL} & 0.8618 & 0.3790 & 0.8039 & 0.5026 \\
  w/o \emph{MLM\&MTM}  & 0.8598 & 0.3815  & 0.8020 & 0.5044  \\
  w/o \emph{MLM\&MTM\&ICL} & 0.8593 & 0.3810 & 0.8008 & 0.5061  \\ \hline
 w/o \emph{Field-level Masking} & 0.8605 & 0.3785 & 0.8054 & 0.5015 \\
w/o \emph{Joint Reconstruction} & 0.8618 & 0.3820 & 0.8050 & 0.5036 \\ 
 \hline
\bottomrule       
\end{tabular}
}
\vspace{-5pt}
\end{table}

Firstly, we evaluate the impact of pretraining objectives (\ie, \emph{MLM}, \emph{MTM}, \emph{ICL}) by eliminating them from the pretraining stage. 
Note that removing all three objectives means that the pretraining stage does not exist. 
The results are reported in Table~\ref{tab:ablation}.
\begin{itemize}[leftmargin=10pt]
    \item As we can observe, the optimal performance is achieved when three losses are deployed simultaneously, and removing each loss will degrade performance, while eliminating all losses results in the lowest performance. These phenomena demonstrate that each component contributes to the final performance.
    \item Removing ICL (w/o \emph{ICL}) obtains better performance than removing MLM\&MTM (w/o \emph{MLM\&MTM}), indicating that joint modeling for MLM\&MTM tasks can learn meaningful cross-modal alignments even without ICL.
\end{itemize}

Next, to further investigate the effect of jointly masked modality modeling, we design the following two variants:
\begin{itemize}[leftmargin=10pt]
    \item \textbf{w/o \emph{Field-level Masking}}: We replace the field-level masking for textual data with the common random token-level masking, as discussed in Section~\ref{sec:method field level data masking}.
    \item \textbf{w/o \emph{Joint Reconstruction}}: The reconstruction of one masked modality depends only on itself and no longer relies on the help from the other modality.
\end{itemize}
The results are shown in Table~\ref{tab:ablation}. 
The performance drops, especially on the Logloss metric, when we either remove the field-level masking strategy or eliminate the joint reconstruction.
Such a phenomenon demonstrates the importance of fine-grained feature-level alignment, which ensures the feature-level communication and interaction between ID-based models and PLMs for dual modalities.

\subsection{Hyperparameter Study}

\begin{figure}
    \centering
    \includegraphics[width=0.44\textwidth]{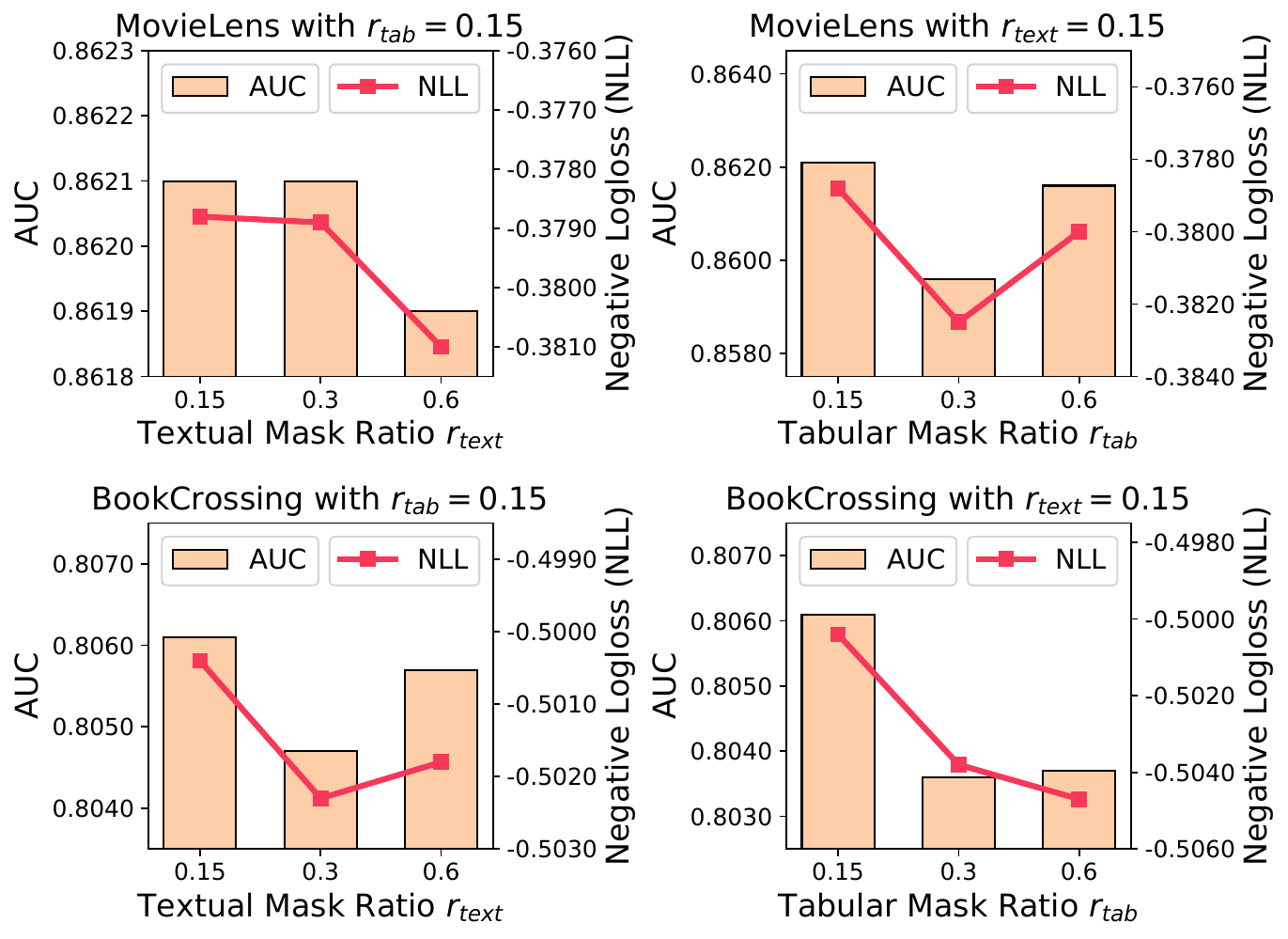}
    \vspace{-10pt}
    \caption{The hyperparameter study on textual mask ratio $r_{text}$ (left column) and tabular mask ratio $r_{tab}$ (right column) on MovieLens-1M (top) and BookCrossing (bottom) datasets.}
    \vspace{-5pt}
    \label{fig:mask_ratio}
\end{figure}

\subsubsection{The Impact of Mask Ratio $r_{text}$ and $r_{tab}$}
Since there are two independent mask ratio for textual and tabular modalities, we fix the mask ratio on the one side and alter the mask ratio on the other side from $\{0.15,0.3,0.6\}$. 
The results are in Figure~\ref{fig:mask_ratio}, from which we observe that the best performance is generally achieved when both mask ratios are relatively small (\ie, 0.15). The reason is that excessive masking might lead to ambiguity in the target modality data, which hurts the model pretraining~\cite{devlin2018bert,liu2019roberta}. So we set the $r_{text}$ and $r_{tab}$ both to 15\%.

\begin{figure}
    \centering
    \includegraphics[width=\linewidth]{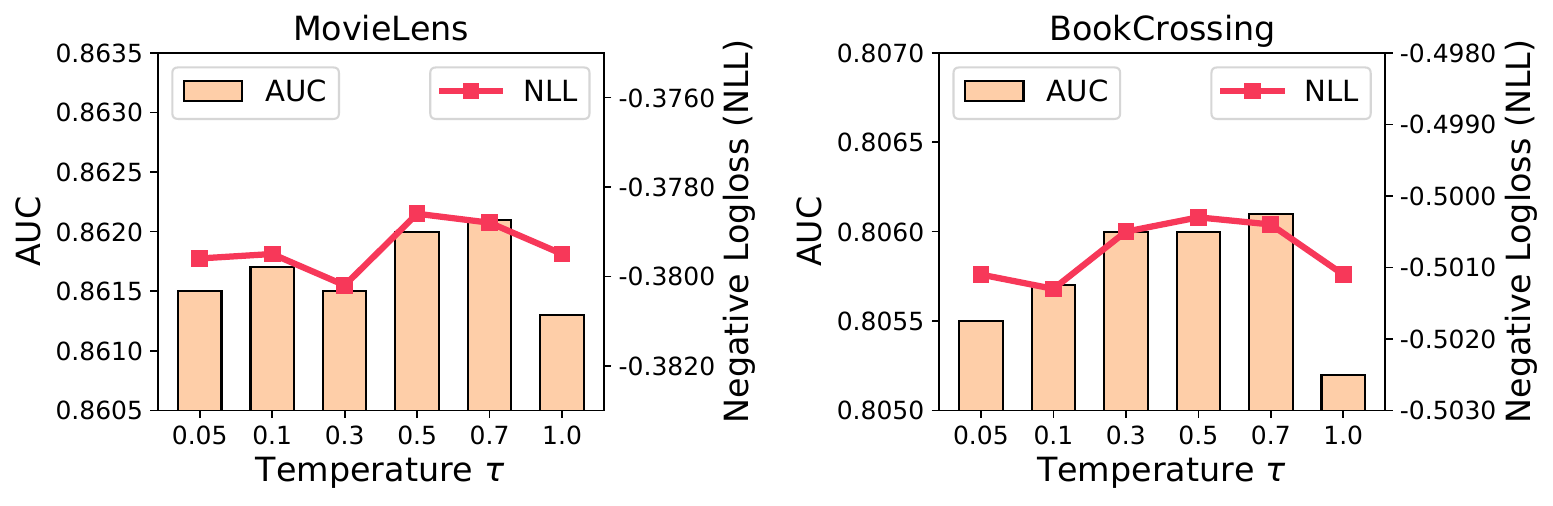}
    \vspace{-20pt}
    \caption{The hyperparameter study on the temperature $\tau$. }
    \vspace{-10pt}
    \label{fig:temperature}
\end{figure}

\subsubsection{The Impact of Temperature Hyperparameter $\tau$}
The temperature $\tau$ controls the sharpness of estimated distribution.
We select $\tau$ from $\{0.05,0.1,0.3,0.5,0.7,1.0\}$, and report the results in Figure~\ref{fig:temperature}. 
When the temperature gradually grows, the performance increases first and then decreases, with the optimal temperature choice between $0.5$ and $0.7$. 
As suggested in previous works~\cite{wang2021understanding,chen2020simple}, a too small temperature would only concentrate on the nearest sample pairs with top scores, and a too large temperature might lead to a flat distribution where all negative sample pairs receive almost the same amount of punishment. 
Both of them will hurt the final performance~\cite{wang2021understanding,li2023ctrl}. Therefore, we choose $\tau=0.7$ in our approach.

\subsection{Analysis on Fine-grained Alignment}
\label{sec:case study}

\subsubsection{Feature-level Alignment}
\begin{figure}
    \centering
    \hfill
    \includegraphics[width=.22\textwidth]{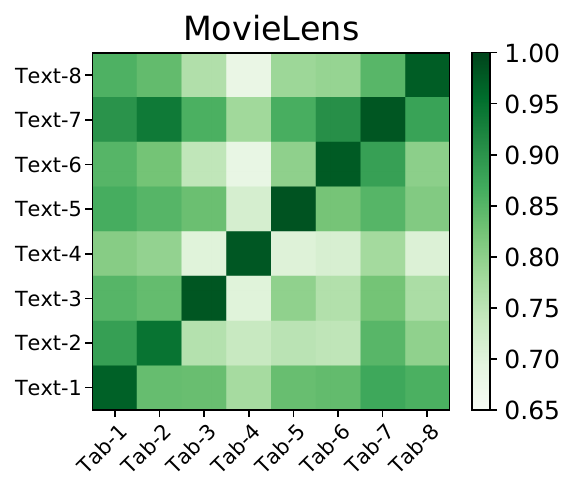}
    \hfill
    \includegraphics[width=.22\textwidth]{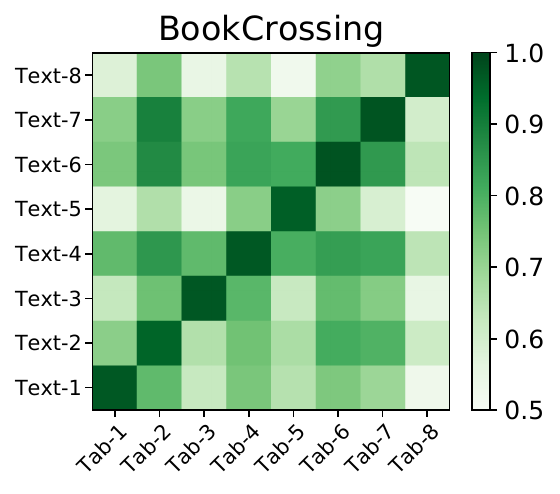}
    \hfill
    \vspace{-10pt}
    \caption{Visualization of similarities between the sample representations of masked textual and tabular data. 
    "Text-$f$" and "Tab-$f$" denote that we mask the $f$-th field of the input data of textual or tabular modalities, respectively.}
    \vspace{-10pt}
    \label{fig:heatmap}
\end{figure}

We conduct case studies to further explore the fine-grained feature-level alignment established during pretraining.
For a dual-modality input pair $(\textbf{x}_i^{text},\textbf{x}_i^{tab})$, we first perform field-level data masking to mask out each field respectively, resulting in $F$ pairs of corrupted inputs $\{(\textbf{x}_{i,(f)}^{text},\textbf{x}_{i,(f)}^{tab})\}_{f=1}^F$, where $\textbf{x}_{i,(f)}^{text}$ and $\textbf{x}_{i,(f)}^{tab}$ denote that we solely mask the $f$-field of the textual or tabular data.
Then we employ the PLM and ID-based model to encode them into $F$ pairs of normalized sample representations $\{(\textbf{z}_{i,(f)}^{text},\textbf{z}_{i,(f)}^{tab})\}_{f=1}^F$.
Next, we compute the mutual similarity scores (measured by dot product) over each cross-modal representation pair, and visualize the heat map in Figure~\ref{fig:heatmap}.

We can observe that the similarity score varies a lot for cross-modal input pairs with different masked fields, and the top-similar score is achieved for pairs with the same masked field (\ie, on the diagonal).
This indicates that FLIP can perceive the changes among field-level features for both modalities, and further maintain a one-to-one feature-level correspondence between the two modalities.

 \begin{figure}
 \vspace{-10pt}
    \subfigure[FLIP$_{\text{w/o }MLM\&MTM}$]{
        \includegraphics[width=0.35\linewidth]{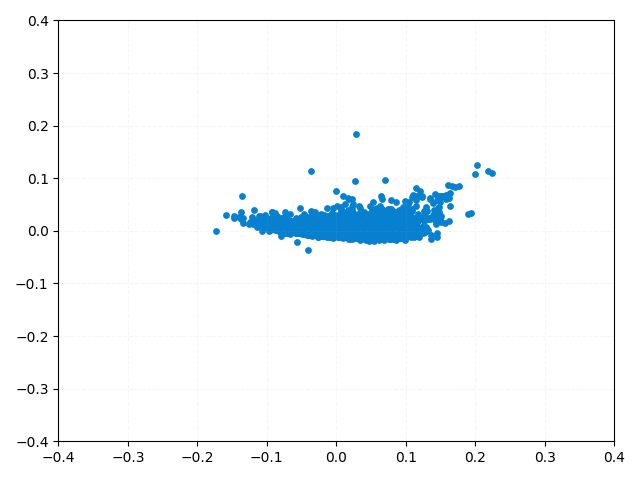}
    }
    \quad
    \subfigure[FLIP$_{\text{w/o }MLM}$]{
        \includegraphics[width=0.35\linewidth]{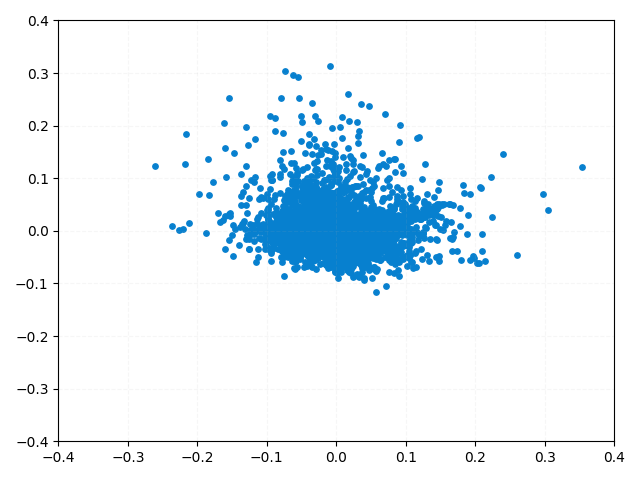}
    }
    
    \subfigure[FLIP$_{\text{w/o }MTM}$]{
        \includegraphics[width=0.35\linewidth]{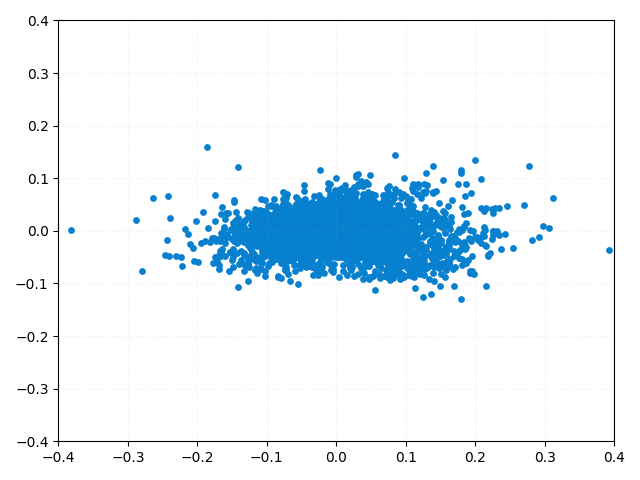}
    }
    \quad
    \subfigure[FLIP]{
        \includegraphics[width=0.35\linewidth]{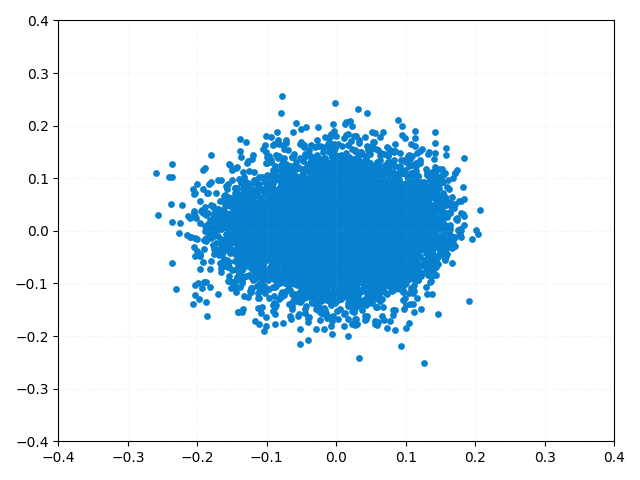}
    }
 \vspace{-10pt}
    \caption{The visualization of feature ID embeddings learned by different model variants on MovieLens-1M. We use SVD to project the feature embedding matrix into 2D data.}
 \vspace{-5pt}
    \label{fig:representation}
\end{figure} 

\subsubsection{Visualization}
we investigate the impact of the fine-grained alignment on the feature ID embedding learning.
Following previous work~\cite{qiu2022contrastive}, we adopt SVD~\cite{koren2008factorization} decomposition to project the learned ID embeddings from the ID-based model into 2D data.
In Figure~\ref{fig:representation}, we visualize the ID embeddings learned by four variants FLIP$_{\text{w/o } MLM\&MTM}$, FLIP$_{\text{w/o } MLM}$, FLIP$_{\text{w/o } MTM}$ and FLIP.
Note that FLIP$_{\text{w/o } MLM\&MTM}$ indicates the instance-level contrastive learning (ICL) variant without the MLM and MTM objectives. 
We have the following observations:
\begin{itemize}[leftmargin=10pt]
    \item The feature ID embeddings learned by FLIP$_{\text{w/o } MLM\&MTM}$ (\ie, ICL) collapse into a narrow cone, suffering from severe representation degeneration problem.
    In contrast, FLIP obtains feature ID embeddings with more distributed latent patterns, thus better capturing the feature diversity.
    This highlights the effectiveness of fine-grained alignment in promoting representation learning and mitigating the representation degeneration.
    \item Comparing FLIP with FLIP$_{\text{w/o } MLM}$ or FLIP$_{\text{w/o } MTM}$, we find that removing either MLM or MTM objective makes the learned embeddings more indistinguishable, demonstrating the necessity of dual alignments between modalities for learning effective representations.
\end{itemize}

\section{Related Work}

\subsection{ID-based Models for CTR Prediction}

CTR prediction serves as a core function module in personalized online services, including online advertising, recommender systems, etc~\cite{zhang2017deep}. ID-based models follow the common design paradigm: embedding layer, feature interaction (FI) layer, and prediction layer. 
These models take as input one-hot features of tabular data modality, and employ various interaction functions to capture collaborative signals among features. 
Due to the significance of FI in CTR prediction, numerous studies focus on designing novel structures for the FI layer to capture more informative and complex feature interactions. Wide\&Deep~\cite{WDL} combines a wide network and a deep network to achieve the advantages of both. DCN~\cite{DCNv1} and DCNv2~\cite{DCNv2} improve Wide\&Deep by replacing the wide part with a cross network to learn explicit high-order feature interactions. DeepFM~\cite{DeepFM} combines DNN and FM, and xDeepFM~\cite{xDeepFM} extends DeepFM by using a compressed interaction network (CIN) to capture feature interactions in a vector-wise way. Furthermore, explicit or implicit interaction operators are designed to improve performance, such as the productive operator~\cite{PNN}, the logarithmic operator~\cite{AFN} and the attention operator~\cite{AFM,AutoInt}.


\subsection{PLMs for CTR Prediction}
\label{sec:related work lm for ctr}

Pretrained Language Models (PLMs) have demonstrated exceptional success in a wide range of tasks, owing to their extensive knowledge and strong reasoning abilities~\cite{radford2018improving,radford2019language,brown2020language,lin2024rella}.
Inspired by these achievements, the application of PLMs for recommender systems has received more attention~\cite{lin2023can,liu2023pre,li2023large,zhu2023large,chen2023large,fan2023recommender,wu2023survey,2023arXiv231112338C,zhu2023large,wang2024towards,du2024disco}. 
Different from one-hot encoding in ID-based models, this line of research needs to convert the raw data into textual modality, thus retaining the original semantic information of features. 

Recent efforts to adapt PLMs for CTR prediction have yielded several breakthroughs~\cite{cui2022m6,li2023text,bao2023tallrec,zhang2023collm,yuan2023go,wang2023zero,ma2023large,hua2023up5,wei2023llmrec}. 
For instance, CTR-BERT~\cite{CTRBERT} leverages a two-tower structure with a user BERT and item BERT for final CTR prediction. 
PTab~\cite{liu2022ptab} pretrains a BERT model by masked language modeling, and then finetunes it on downstream tasks. 
P5~\cite{P5} uniformly converts different recommendation tasks into text generation tasks with T5 backbone~\cite{T5}. 
However, PLMs encounter challenges in capturing field-wise collaborative signals and discerning features with subtle textual differences.
Recent works~\cite{yuan2023go,li2023ctrl,ren2023representation} have tried to address this problem by adding text features into the ID-based model or integrating ID information into the PLM. However, they either overlook cross-modal interactions or depend solely on coarse-grained ICL, which is insufficient to capture fine-grained feature-level modality interactions.


Therefore, we propose to conduct fine-grained feature-level alignment between ID-based models and PLMs via the jointly masked tabular/language modeling, which learns fine-grained interactions between tabular IDs and word tokens by the way of mask-and-predict.
In the finetuning stage, we also propose to jointly tune both models, and thus leverage the benefits of both textual and tabular modalities to achieve superior CTR prediction performance.
\section{Conclusion}
In this paper, we propose FLIP, a model-agnostic framework that achieves the fine-grained alignment between ID-based models and PLMs. 
We view tabular data and transformed textual data as dual modalities, and design a novel fine-grained modality alignment pretraining task. Specifically, the joint reconstruction for masked language/tabular modeling (with specially designed masking strategies) and cross-modal contrastive learning are employed to accomplish feature-level and instance-level alignments, respectively. 
Furthermore, we propose to jointly finetune the ID-based model and PLM to achieve superior performance by adaptively combining the outputs of both models.
Extensive experiments show that FLIP outperforms state-of-the-art baselines on three real-world datasets, and is highly compatible with various ID-based models and PLMs.


\begin{acks}
The Shanghai Jiao Tong University team is partially supported by National Natural Science Foundation of China (62177033, 62076161) and Shanghai Municipal Science and Technology Major Project (2021SHZDZX0102). The work is also sponsored by Huawei Innovation Research Program. We thank MindSpore~\cite{mindspore} for the partial support of this work.
\end{acks}


\bibliographystyle{ACM-Reference-Format}
\bibliography{acmart}

\end{document}